\documentclass[prd,aps,showpacs,floats,letterpaper,floatfix,
twocolumn,nofootinbib]{revtex4}
\usepackage{afterpage}
\usepackage{amssymb,amsmath,amsbsy}
\usepackage[dvips]{graphicx}
\usepackage{dcolumn}
\usepackage[usenames]{color}
\usepackage{euscript}
\usepackage{mathrsfs}
\usepackage{enumerate}
\begin{document}
\title{
A generalization of Ryan's theorem:
probing tidal coupling with gravitational waves
from nearly circular, nearly equatorial, extreme-mass-ratio inspirals
}
\author{Chao Li}\author{Geoffrey Lovelace}
\affiliation{Theoretical Astrophysics, California
Institute of Technology, Pasadena, California 91125}

\begin{abstract}
Extreme-mass-ratio inspirals (EMRIs) and intermediate-mass-ratio
inspirals (IMRIs)---binaries in which a stellar-mass object
spirals into a massive black hole or other massive, compact body---are
important sources of gravitational waves for LISA and LIGO, respectively.
Thorne has speculated that the waves from EMRIs and IMRIs encode,
in principle, all the details of (i) the central body's spacetime
geometry (metric), (ii) the tidal coupling (energy and angular
momentum exchange) between the central body and orbiting object, and
(iii) the evolving orbital elements. Fintan Ryan has given a first partial
proof that this speculation is correct: Restricting himself to
nearly circular, nearly equatorial orbits and ignoring tidal coupling,
Ryan proved that the central body's metric is encoded in the waves.
In this paper we generalize Ryan's theorem.  Retaining Ryan's restriction
to nearly circular and nearly equatorial orbits, and dropping the
assumption of no tidal coupling, we prove that Thorne's conjecture
is nearly fully correct: the waves encode not only the central body's
metric but also the evolving orbital elements and (in a sense slightly
different from Thorne's conjecture) the evolving tidal coupling.
\end{abstract}


\date{\today}

\maketitle


\section{Introduction and Summary}
\label{sec:introduction}

The LIGO-GEO-VIRGO-TAMA network of broadband
ground-based laser interferometers, aimed
at detecting gravitational waves in the high-frequency band $10$--$10^4$
Hz, is already operating at or near
its
initial design sensitivities.
In the next decade,
LISA (the Laser Interferometer Space Antenna)
will open up the low-frequency
gravitational-wave window ($10^{-4}$--0.1 Hz).

Among the most important sources of gravitational waves for
LISA
are
extreme-mass-ratio
inspirals (EMRIs), which are systems in which a small object (with
mass \(\mu\sim M_\odot\)) orbits a supermassive black hole or other
central body (boson star \cite{Colpietal1986}\cite{Ryan1997} or soliton star 
\cite{Lee1987} or
naked singularity or ...) with
mass \(M\sim 10^6 M_\odot\). Recently, Brown and collaborators
\cite{brownetal} have estimated that
advanced detectors in LIGO (the Laser Interferometric
Gravitational-Wave Observatory)
may detect up to \(\sim 10 - 30 \text{yr}^{-1}\)
\emph{intermediate
mass ratio inspirals} (IMRIs), which are analogous to EMRIs but have less
massive central bodies
(masses
\(M\) in the range of \(\sim 10^2 - 10^4
M_\odot\)).

Thorne has conjectured\footnote{Thorne's conjecture has grown over
time.  Originally, in the early 1990s, he conjectured (or, more precisely,
asserted!) that the waves encode ``a portion''
of the spacetime geometry (e.g.\  p.\ 326 of \cite{Abramovici1992}). By 1994,
when Fintan Ryan proved his theorem, Thorne was arguing that the entire 
spacetime 
geometry would be encoded (see, e.g., the introduction to 
Ryan's paper \cite{ryanThm}). In 2002, when thinking about how LISA
might test the laws of black-hole physics, Thorne realized that the tidal 
coupling might also
be encoded along with the central body's spacetime geometry; 
see Ref.\ \cite{Hawkings60th}.  Only recently, when advising the
authors about their research, did Thorne realize that the evolving orbital 
elements might
also be extractable (private communication).}
that the waves from an EMRI or IMRI
contain, encoded in themselves (at least in principle):
(i) the spacetime geometry (metric)
of the massive central body, (ii) the tidal coupling (evolving
rate of energy and angular momentum exchange) between the orbiting object
and the central body, and (iii) the evolving orbital elements.
This conjecture (which has been partially proved; see below) has
motivated placing EMRIs high on LISA's list of target sources
\cite{LISAsciencegoals}, and has motivated research to:
(a) prove Thorne's conjecture with the widest generality possible, or,
if it is false, determine what information actually \emph{is} encoded in the
EMRI and IMRI waves \cite{brownetal}\cite{chaoEccentricOrbits};
(b) develop data analysis techniques for searching for EMRI and IMRI
waves in LISA \cite{FirstCutLDA}\cite{GairWen} and LIGO \cite{DuncanInPrep} 
data; 
(c) scope
out the accuracy with which LISA and LIGO can extract the encoded
information from EMRI and IMRI waves (and if the central body appears to
be a black hole, the accuracy with which its properties agree with those of
a hole)~\cite{CollinsAndHughes}\cite{GlampedakisAndBabak}; and (d) develop data
analysis techniques for extracting the waves' information 
\cite{StroeerGairVecchio}.

Fintan Ryan \cite{ryanThm} has
proved a theorem that is an important step toward verifying Thorne's
conjecture.  Specifically, he has proved that it is possible in principle to
recover the full spacetime geometry from EMRI waves under the following 
assumptions:
i) the central body is general-relativistic, stationary, axisymmetric,
reflection-symmetric, and asymptotically-flat (SARSAF),
ii) the small object travels on
a nearly circular and nearly equatorial orbit, and iii) there is no
tidal coupling. Moreover, Ryan has shown that the multipole moments that
determine the spacetime geometry are
\emph{redundantly} encoded in the gravitational waves and can be extracted
using \emph{either} of the two precession frequencies (about a circular orbit 
and
about the equatorial plane) \emph{or} the waves' phase evolution.

The purpose of this paper is to generalize Ryan's theorem.
We retain assumptions (i) and (ii)
(SARSAF spacetime and nearly circular, nearly equatorial orbit)
but relax
assumption (iii) by allowing for a small amount of tidal coupling. 
We show that in this case, Thorne's conjecture is nearly correct:
the waves encode not only the central body's
metric but also the evolving orbital elements and (in a sense slightly
different from Thorne's conjecture) the evolving tidal coupling.
(Assumption (ii), that the orbit is nearly circular and nearly
equatorial, is relaxed in a companion paper by Li
\cite{chaoEccentricOrbits}).

Motivated by the result of Fang and
Lovelace \cite{huaGeoffreyTides} that
the only unambiguous part of the tidal coupling is the
time-dependent,
dissipative portion (at least when the
central body is a non-spinning black hole and the orbit is large
and circular),
we characterize the
tidal coupling by the
rates of energy and angular momentum exchange between the central
body and the orbiting object, \(\dot{E}_\text{body}\)
and \(\dot{L}_\text{body}\). (Throughout
this paper, a dot means derivative with respect to the coordinate
time $t$, which is the time measured by an inertial
observer in the asymptotically flat region of the spacetime.)
Actually, we only need to consider \(\dot{E}_\text{body}\), because
once it is known, \(\dot{L}_\text{body}\) can be deduced from the
standard energy-angular momentum relation for circular orbits and their
influence on waves and tides, \(\dot E = \Omega_\text{orbit} \dot L \).
(Here \(\Omega_\text{orbit}\) is the orbital angular velocity, which
is the same as the waves' observed primary angular frequency aside
from a factor 2.)

This paper is organized as follows: In Sec.~\ref{sec:spacetimegeometry},
we begin by noting that,
when there is a small amount of tidal coupling
(as we assume), then
the redundancy in
Ryan's analysis is broken.
One can still
use
Ryan's algorithm for the precession frequencies
to recover the central body's
spacetime geometry. Then,
by making use of the observed (time-independent) spacetime
geometry and the measured, evolving amplitudes associated with the precession
frequencies, one can also recover from the EMRI waves the evolving
orbital parameters.  Having relied on non-dissipative aspects of the waves
to deduce the spacetime geometry and orbit, one can then---as we show in
Sec.~\ref{sec:tidalCoupling}---use the waves' dissipation-induced phase 
evolution
to deduce the tidal coupling.

In our somewhat delicate discussion of deducing the tidal coupling
(Sec.~\ref{sec:tidalCoupling} ), we begin by noting
that the sum of the power radiated to infinity and the power fed into
the central body via tidal coupling,
\(
\dot E_\text{total} = \dot E_\infty + \dot E_\text{body}
\)
is equal to the power lost from the orbit, which
can be deduced from the waves' observed phase evolution.  The
central body influences this observed \(\dot E_\text{total}\)
in two ways: (i) by generating a nonzero \(\dot E_\text{body}\), the quantity
that interests us, and (ii) by very slightly altering \(\dot E_\infty\).
To help quantify these two body influences, in Sec.\ \ref{subsec:TCPE} we
show how one can deduce, from the observations, the rate
\(\dot E_{\infty\text{NBI}}\) that energy
would be radiated to infinity if there were \emph{no body influences}.  
The difference 
between the measured
\(\dot E_\text{total}\) and the deduced \(\dot E_{\infty\text{NBI}}\)
is the influence of the body's structure on
the total energy loss from the orbit,
\( \dot E_\text{total,BI} \equiv \dot E_\text{total} 
- \dot E_{\infty\text{NBI}}\).
This measured/deduced body influence on the total energy loss consists of two
tiny pieces:
the power that actually goes into the body via tidal coupling,
\(\dot E_\text{body}\), and the body's tiny influence on the power
radiated to infinity, \(\dot E_{\infty\text{BI}} \equiv \dot E_\infty
- \dot E_{\infty\text{NBI}}\):
\begin{equation}
\dot E_\text{total,BI} \equiv \dot E_{\infty\text{BI}} + \dot E_\text{body}\;.
\end{equation}

In principle (as described above), from the observational data plus 
general-relativity
theory we know the body's influence on the total energy loss
\(\dot E_\text{total,BI}\)
with complete precision.  This is not quite what Thorne conjectured,
but it is close, and it is the only complete-precision statement we have
been able to make about measuring the influence of tidal coupling.

Thorne conjectured we could deduce \(\dot E_\text{body}\)
from the observed waves.  This, in fact appears not to be possible
(in principle) with complete precision.  However, we argue in
Sec.\ \ref{sec:ambiguity} and the Appendix that, if the central body
is highly compact, then
the unknown \(\dot E_{\infty\text{BI}}\) will be smaller than
\(\dot E_\text{body}\) by \(\sim v^n \ll 1\), where
\(v\) is the orbital velocity and \(n\) is some high power; and we
show that, when the body's external metric is that of Schwarzschild
or Kerr, then \(n=5\).  As a result, aside from a very small
O(\(v^n)\) uncertainty due to the influence of the body on the energy
radiated to infinity, the tidal coupling power \(\dot E_\text{body}\)
is equal to the known influence of the body on the total
energy loss \(\dot E_\text{total,BI}\).

A brief conclusion is made in Sec.~\ref{sec:conclusion}.

\section{Extracting the Spacetime geometry and orbital elements}
\label{sec:spacetimegeometry}\label{sec:spacetimeGeometry}

Aside from allowing tidal coupling, we treat the same class of EMRIs
as did Ryan:

\emph{First,} we assume the central body's exterior spacetime
is a vacuum, stationary, axisymmetric, reflection symmetric, and
and asymptotic flat (SARSAF) solution of Einstein's equations. The exterior 
spacetime 
metric can be written as (e.g., Eq.~(7.1.22) of Ref.~\cite{wald})
\begin{eqnarray} ds^2=-F(dt-\omega
d\phi)^2+\frac{1}{F}[e^{2\gamma}(d\rho^2+dz^2)+\rho^2d\phi^2],\end{eqnarray}
where $F,\omega$ and $\gamma$ are functions of $\rho$ and $|z|$.
In SARSAF spacetimes, there is a
one to one correspondence between the spacetime metric and a
series of scalar multipole moments $(M_{2i},S_{2i+1}),\quad
i=0,1,\cdots$ \cite{multipole1,multipole2}. Here \(M_0\equiv M\)
is the mass of the central body, \(S_1\) is its spin, \(M_2\) is
its mass quadrupole moment, etc. To extract the geometry of the
spacetime surrounding the central body, it is sufficient to
extract the multipole moments \(\left\{M_\ell,S_\ell\right\}\)
\cite{ryanThm}.

\emph{Second,} we let a small object with mass \(\mu \ll M\) move about the
central body in a nearly circular,
nearly equatorial orbit.

For precisely circular, equatorial, geodesic motion, the waves obviously have
a single fundamental frequency \(\Omega_\phi\) that is associated with
the circular motion \(\phi = \Omega_\phi t\).  When the geodesic orbit is
slightly nonradial, it is easy to show that its radius \(\rho\)
undergoes periodic motion with some angular frequency \(\Omega_\rho\);
and when slightly nonequatorial, its vertical coordinate \(z\) undergoes
periodic motion with another angular frequency \(\Omega_z\).
These geodesic motions give rise to gravitational waves that are
triperiodic: a discrete spectrum with frequencies equal to
\(\Omega_\phi\), \(\Omega_\rho\), \(\Omega_z\), and their sums
and differences (see Ref.\ \cite{chaoEccentricOrbits} for a proof, patterned
after the proof by Drasco and Hughes \cite{Kerrgeodesics} for the Kerr metric).  
The
frequency difference \(\Omega_\rho - \Omega_\phi\) shows up as an
orbital periapsis precession, and \(\Omega_z - \Omega_\phi\)
as an orbital plane precession; these precessions produce
corresponding modulations of the gravitational waveforms.

In our case, the orbits are not geodesics; they evolve due
to gravitational radiation reaction.
Because of the extreme mass ratio,
the radiation
reaction
can be described by the adiabatic
approximation. In this approximation, on the timescale of an
orbital
period,
the small object moves
very nearly
along a
geodesic of the central body's gravitational field.
On a timescale much larger than the orbital period,
the object moves from one geodesic to another as it
loses energy and angular momentum to gravitational radiation. It follows that
the three frequencies
\(\left\{\Omega_\phi(t),\Omega_\rho(t),\Omega_z(t)\right\}\) each evolve
with time on the radiation reaction timescale which is much longer than
the orbital periods.

In principle, a large amount of information can be encoded in the
time evolution of the waves' three fundamental frequencies
\(\left\{\Omega_\phi(t), \Omega_\rho(t), \Omega_z(t)\right\}\) and
the complex amplitudes (amplitudes and phases) of the various
spectral components.  The largest amplitudes are likely to be those
for the second harmonic of \(\Omega_\phi\) and for the two precessions,
\(h_{2\Omega_\phi}(t)\), \(h_{\Omega_\rho - \Omega_\phi}(t)\), and
\(h_{\Omega_\rho - \Omega_z}(t)\).  We shall call these the
primary-frequency component, and the precessional components of
the waves.  To simplify our prose, we shall refer to $\Omega_\rho$
and $\Omega_z$ as the ``precession frequencies'' even though the
actual frequencies of precession are $\Omega_\rho - \Omega_\phi$
and $\Omega_z - \Omega_\phi$.

Thorne's conjecture can be expressed mathematically as the claim that
these time evolving
frequencies and amplitudes encode
fully and separably,
\begin{enumerate}
\item the values of all the central body's
multipole moments $\{M_\ell, S_\ell\}$,
\item the rates $\dot{E}_{\rm body}$ and
$\dot{L}_{\rm body}$ at which the
orbiting object's tidal pull deposits energy and angular momentum
into the central body, and
\item the time-evolving orbital elements,
i.e. the
orbit's
semi-latus rectum $p(t)$, eccentricity $e(t)$ and inclination
angle $\iota(t)$.
\end{enumerate}

Ryan's theorem\ \cite{ryanThm} states that, if there is no tidal coupling, then
all the SARSAF moments $\{M_{2i}, S_{2i+1}\}$
are encoded in the time evolving frequencies fully, separably, and
redundantly.
Ryan did not explicitly address the encoding of the three orbital elements
$p(t)$, $e(t)$ and $\iota(t)$. However, their
encoding is an almost trivial extension of his analysis:

Specifically, Ryan noticed that the three fundamental frequencies
are independent of $e$ and $\iota$ to first order in these small
quantities, i.e., they are functions solely of the moments and the
semi-latus rectum $p$. One can eliminate $p$ by regarding the
precession frequencies $\Omega_z$ and $\Omega_\rho$ as
functions of the moments and $\Omega_\phi$, or equivalently as
functions of the moments and the Post-Newtonian (PN) expansion
parameter $v\equiv (M \Omega_\phi)^{1/3} \simeq ($orbital
velocity). Expanding $\Omega_z(v; S_\ell, M_\ell)$ and
$\Omega_\rho(v; S_\ell, M_\ell)$ in powers of $v$, Ryan found the
following pattern of coefficients (with each moment first
appearing at a different power of $v$), from which all the
moments can be extracted separably (Eqs.~(18) -- (19) of
Ref.~\cite{ryanThm}):\begin{eqnarray}
\frac{\Omega_\rho}{\Omega_\phi}&=&3v^2-4\frac{S_1}{M^2}v^3+\left(\frac{9}{2}
-\frac{3M_2}{2M^3}\label{eq:precessionFreqs}
\right)v^4+\cdots\nonumber\\
\frac{\Omega_z}{\Omega_\phi}&=&2\frac{S_1}{M^2}v^3+\frac{3M_2}{2M^3}v^4+\cdots
\label{eq:precessionfrequencies}. \end{eqnarray}

This result leads to Ryan's algorithm
for extracting information.
\emph{First,} from
the waves' observed time-evolving
precession
frequencies and
time-evolving primary
frequency, one can deduce the functions
$\Omega_{z,\rho}(\Omega_\phi)$ and thence $\Omega_{z,\rho}(v)$;
\emph{second,}
expanding in powers of $v$, one can then read out the multipole moments
\(\left\{M_\ell,S_\ell\right\}\)
from either \(\Omega_z(v)\) or \(\Omega_\rho(v)\).

We almost trivially augment onto Ryan's algorithm the following
steps for extracting the time-evolving orbital elements:
\emph{Third,} knowing the moments and thence
the metric, one can use the geodesic equation to deduce $p(t)$ from
$\Omega_\phi(t)$.
\emph{Fourth},
one can use wave-generation theory
and knowledge of the metric to
deduce $e(t)$ and \(\iota(t)\) from the amplitudes
$h_{\Omega_\rho-\Omega_\phi}$ and $h_{\Omega_z - \Omega_\phi}$
of the wave modulations due to
periapse precession and orbital plane precession.

\section{Probing Tidal Coupling}
\label{sec:tidalCoupling}
We now drop Ryan's restriction of no tidal coupling. This does not
alter
Eqs.~\eqref{eq:precessionfrequencies} for \(\Omega_\rho\) and
\(\Omega_z\) as functions of $v$, i.e. of
the orbital frequency \(\Omega_\phi\),
since all
three frequencies only depend on the geodesic motion and hence
only depend on the multipole moments \(\left\{M_\ell,S_\ell\right\}\).
On the other hand, the evolution
of the frequencies,
\emph{as functions of time, will} depend on the tidal
coupling.

More generally, we can divide the physical quantities
of our analysis
into two
categories: i) ``static:'' those quantities related to the geodesic motion of 
the
orbiting object, and ii) ``dynamic:'' those quantities related to the
inspiral of the object (i.e., to the
evolving rate at which the object
moves from
geodesic to geodesic). All static quantities are independent of
tidal coupling and all dynamic quantities
depend on it.

This suggests that Ryan's analysis can be extended to include
tidal coupling. First, the
static quantities can be used to deduce the the
central body's multipole moments, just as in Ryan's original argument
as sketched above.
Then, the dynamic quantities, combined with knowledge of the spacetime
metric, can be used
to extract tidal-coupling information.
This extension is discussed
in the following subsections.

\subsection{The phase evolution when tidal coupling is neglected}
Following Ryan, we characterize the phase evolution of EMRI waves by
the number of primary-frequency cycles of waves per logarithmic frequency
interval, as a function of the primary waves' slowly increasing
frequency $f = \Omega_\phi/\pi$. This quantity can be written as
(Eq.~(4) of Ref.~\cite{ryanThm})
\begin{eqnarray}\label{eq:phaseDef}
\Delta N(f) \equiv \frac{fdt}{d\ln f} = \frac{f^2}{df/dt}.
\end{eqnarray}
This
phase evolution $\Delta N(f)$ can be measured by gravitational-wave detectors
with high precision.

If there is no tidal coupling \emph{and} no other influence of the structure of
the central body on the waves, as Ryan assumed, then
it is possible to
read off the multipole moments (and also the small object's mass\footnote{The 
mass of 
the small object can be determined from \(\Delta N(f)\) \emph{even when there is 
tidal coupling}. The leading-PN-order part of the energy flux (equivalently, the 
leading-PN-order part of \(\Delta N(f)\)) is independent of tidal coupling. 
One can thus equate the leading-PN-order parts of \(\Delta N(f)\) and 
\(\Delta N_\text{NBI}\) [Eq.~(\ref{DNNBI})]. After inserting the mass \(M\) 
(obtained from one of the precession frequencies), one can solve for \(\mu\). 
The precession frequencies, in contrast, are independent of \(\mu\) 
[Eqs.~(\ref{eq:precessionfrequencies})].})
from a PN expansion of $\Delta N(f)$
(Eq.~(57) of
Ref.~\cite{ryanThm}):
\begin{eqnarray}
\Delta N_\text{NBI} =
\frac{5}{96\pi}\left(\frac{M}{\mu}\right)v^{-5}\Bigg[1+
\frac{743}{336} v^2
-  4 \pi |v|^3 & & \nonumber\\ 
+ \frac{113}{12}\frac{S_1}{M^2} v^3
+ \left(\frac{3058673}{1016064}-\frac{1}{16}\frac{S_1^2}{M^4}
+5\frac{M_2}{M^3}\right) v^4 & & \nonumber\\
+ \sum_{\ell=4,6,...}\frac{(-1)^{\ell/2}(4\ell+2)(\ell+1)!!
\left[M_\ell\text{ + TNILM}\right] v^{2\ell}}
{3 \ell!! M^{\ell+1}} & & \nonumber\\
+ \sum_{\ell=3,5,...}\frac{(-1)^{(\ell-1)/2}(8\ell+20)\ell!! 
\left[S_\ell\text{ + TNILM}\right] v^{2\ell+1} }
{3 (\ell-1)!! M^{\ell+1}} \Bigg]. & & \nonumber\\
\label{DNNBI}
\end{eqnarray}
Here ``NBI'' stands for \emph{no body influence} and ``TNILM'' stands for
\emph{terms nonlinear in lower moments}.
[Recall that $v= (M\Omega_\phi)^{1/3} = (\pi M f)^{1/3}$.]
So long as tidal coupling is negligible, then,
the spacetime multipole
moments
can be determined
redundantly from \emph{either} \(\Delta N(f)\) [Eq.\ (\ref{DNNBI})]
\emph{or}
the periapse precession frequency \(\Omega_{\rho}(\Omega_\phi)\)
\emph{or} the orbital-plane precession frequency \(\Omega_{z}(\Omega_\phi)\)
[Eqs.\ (\ref{eq:precessionfrequencies})].

\subsection{Tidal coupling and the phase evolution}
\label{subsec:TCPE}
When tidal coupling effects
are included, the redundancy is broken. The multipole moments
\(\{M_\ell,S_\ell\}\) can still [Eq.~(\ref{eq:precessionFreqs})] be
determined from \(\Omega_{\rho,z}(\Omega_\phi)\), while
(as the following discussion shows),
the tidal coupling can be determined from
\(\{M_\ell,S_\ell\}\) and \(\Delta N(f)\).


As a preliminary to discussing this, we explain why it is sufficient,
in analyzing tidal coupling, to focus on energy exchange between the
orbit, the body and the waves, and ignore angular momentum exchange.
Since the body is in a (nearly) circular, geodesic orbit, changes in its
orbital energy and angular momentum are related by
\begin{subequations}
\begin{eqnarray} \dot{E}_{\rm orbit} = \Omega_{\phi} \, \dot{L}_{\rm
orbit},
\label{eq:ELorbit}
\end{eqnarray}
aside from second-order corrections due to the slight orbital
ellipticity and inclination angle. Our entire analysis is restricted to
first-order corrections, so those second-order corrections are negligible. 
Similarly,
since the energy and angular momentum radiated to infinity are carried
by the primary waves, with angular frequency $\omega = 2\pi f = 2
\Omega_\phi$ (aside from negligible contributions from the precessions,
which are second order in the ellipticity and inclination), each
graviton carries an energy $\hbar\omega = 2\hbar\Omega_\phi$ and
an angular momentum $2\hbar$ (with this last 2 being the graviton spin).
Therefore,
the energy and angular momentum radiated to infinity are
related by
\begin{eqnarray} \dot{E}_\infty = \Omega_{\phi} \, \dot{L}_\infty.
\label{eq:ELinfty}
\end{eqnarray}
Conservation of energy and of angular momentum, together with Eqs.\
(\ref{eq:ELorbit}) and (\ref{eq:ELinfty}),
 then imply that
\begin{eqnarray} \dot{E}_{\rm body} = \Omega_{\phi}\,  \dot{L}_{\rm
body},\end{eqnarray} \label{EOmegaL}
\end{subequations} for the energy and angular momentum deposited in the body 
by tidal 
coupling.
Equations (\ref{EOmegaL}) imply that, once we understand, observationally, 
the
energy exchange, an understanding of the angular momentum exchange will
follow immediately.

Now turn to the influence of the body's internal structure on the observed
energy exchange.

The total rate that energy is lost from the orbit (which then goes to infinity 
and the 
body)
is related to the phase evolution $\Delta N(f)$ by
\begin{eqnarray}
\dot E_\text{total} = -
\dot{E}_\text{orbit}= - \frac{dE_\text{orbit}}{df}\frac{df}{dt}=
-f^2\frac{dE_\text{orbit}}{df}\frac{1}{\Delta N}.
\label{dotEtotal}
\end{eqnarray}
The phase evolution
$\Delta N$ and the primary frequency \(f\) are known from observation,
and, after using the precession frequencies to compute
the spacetime metric (Sec. \ref{sec:spacetimegeometry}), it is possible
to compute \(dE_\text{orbit}/df\) via the geodesic equation\footnote{
To do this, first insert the multipole moments \(\{M_\ell,S_\ell\}\) into
the geodesic equation. Then, solve the geodesic equation
for the family of circular, equatorial orbits about the central body.
Each orbit \(i\) will have a particular value of
energy \(E_{\text{orbit},i}\) and frequency \(f_i\);
this one-to-one mapping between \(E_\text{orbit}\) and \(f\) can then be used
to compute $dE_\text{orbit}/df$.}. Thus everything on the
right-hand side of Eq.~(\ref{dotEtotal}) can be determined
from observed quantities,
which means that $\dot E_\text{total}$ is measurable.

Another measurable quantity, we claim, is the rate that energy would be lost 
from
the orbit if the body's structure had no influence.  This quantity is 
[by analogy with
Eq.\ (\ref{dotEtotal})]
\begin{eqnarray}
\dot E_\text{total,NBI} =
- f^2\frac{dE_\text{orbit}}{df}\frac{1}{\Delta N_\text{NBI}}.
\end{eqnarray}
Knowing the moments as a function of frequency from measurements of the
precessions, ${\Delta N_\text{NBI}}$ can be computed from the moments via 
Ryan's 
phasing relation\footnote{Ryan calculates the phasing relation to 2PN order 
(i.e., to \(O(v^4)\) past leading order). By extending Ryan's calculation to 
higher post-Newtonian orders, the terms omitted from 
Eq.~(\ref{DNNBI}) can be 
written explicitly.}
(\ref{DNNBI}), and, as we have seen, $dE_\text{orbit}/df$ can also be computed 
from
the observations; so $\dot E_\text{total,NBI}$ is, indeed, observable,
as claimed.  Therefore the
influence of the body's structure on the orbit's total energy loss
\begin{eqnarray}
\dot E_\text{total,BI} = \dot E_\text{total} - \dot E_\text{total,NBI}
\end{eqnarray}
is also observable.

This body influence on the total energy loss consists of two parts: the energy
that goes into the body via tidal coupling, $\dot E_\text{body}$, and a tiny 
body-influenced
modification of the rate that the waves carry energy to infinity
\begin{eqnarray}
\dot E_\text{total,BI}  = \dot E_\text{body} + \dot E_{\infty\text{BI}}\;,
\end{eqnarray}
where
\begin{eqnarray}
\dot E_{\infty\text{BI}} = \dot E_\infty - \dot E_\text{total,NBI}\;.
\end{eqnarray}

Thorne conjectured that the energy exchange due to tidal coupling,
$\dot E_\text{body}$, would be observable.  We doubt very much that it is,
since in general we see no way to determine the relative contributions
of $\dot E_\text{body}$ and $\dot E_{\infty\text{BI}}$ to the observed
total body influence $\dot E_\text{total,BI}$.  The best one can do, in general,
in validating Thorne's conjecture, is to extract the central body's total
influence on the orbital energy loss, $\dot E_\text{total,BI}$.
However, in the special case of a body that is exceedingly compact, we
can do better, as we shall explain in the next subsection.

\subsection{The dependence of \(\dot{E}_\infty\) on the central
body's internal structure}
\label{sec:ambiguity}
%
Consider a central body sufficiently compact that gravity near its surface
blue shifts the orbiting object's tidal field, making it appear like
ingoing gravitational waves as seen by stationary observers.
 This is the case, for example, when the
central body is a black hole. Then, we claim, the ratio
$\dot E_{\infty\text{BI}}/\dot E_\text{body}$ is very small:
\begin{eqnarray}
\frac{\dot E_{\infty\text{BI}}}{\dot E_\text{body}} \sim v^n \ll 1\;,
\label{eq:Ratiovn}
\end{eqnarray}
where $n$ is a large number, very likely 5. For LISA, almost all of
the wave cycles used in extracting information from the waves will be
from radii where $v \lesssim 0.5$ so $v^5 \lesssim 0.03$.  For example,
for a Kerr black hole, if the spin parameter is $a/M\lesssim 0.5$, then
at the innermost stable circular orbit, $v\lesssim 0.5$.  Consequently,
almost all of the measured $\dot E_\text{total,BI}$ will go into the body
itself via tidal coupling, so $\dot E_\text{body}$ will be measured to
good accuracy.

To understand our claim that $\dot E_{\infty\text{BI}}/\dot E_\text{body}
\sim v^n$ for some large $n$, consider a central body whose external
metric is that of a Kerr black hole.  In this case, one can use the
Teukolsky formalism \cite{tpt} (first-order perturbation theory in the
mass ratio $\mu/M$) to compute the energies radiated to infinity and
tidally coupled into the central body.  We have carried out that Teukolsky
analysis for general $a/M$ and present the details for the special case
$a=0$ in the Appendix.  Here we explain the underlying physics.  We
begin with some preliminaries:

We need only consider the primary-frequency waves,
$f=\Omega_\phi/\pi$, since
they account for all the energy loss and transfer, up to
corrections second order in the eccentricity $e$ and inclination
angle $\iota$. This means, correspondingly, that we can restrict
ourselves to a precisely circular and equatorial orbit. The waves
and tidal coupling then have predominantly spheroidal harmonic
order $\ell = m = 2$ and frequency $f$ (angular frequency $\omega
= 2\pi f = 2\Omega_\phi$).  Since we only want to know, to within
factors of order unity, the ratio $\dot E_{\infty\text{BI}}/ \dot
E_\text{body}$, it will be sufficient to restrict ourselves to
these dominant $\ell = m = 2,$ $\omega = 2\Omega_\phi$
perturbations.

In the Teukolsky formalism, these perturbations are embodied in a radial
``wave function'' that can be normalized in a variety of different ways.
The usual normalization, based on the Newman-Penrose field $\psi_4$,
is
bad for physical insight because it treats outgoing waves and ingoing waves
quite differently; see the Appendix.  One normalization that treats them on
the same footing sets the
radial wave function equal
to that of the tidal gravitational field
(``electric-type'' components of the Weyl or Riemann curvature tensor)
measured by ``zero-angular-momentum'' observers, ZAMOs (a family of
observers, each of whom resides at fixed radius $r$ and polar angle
$\theta$).  We shall denote that tidal field [with $e^{-i\omega t}
\times ($spheroidal harmonic) factored out so the field is complex, not
real] by $\mathcal E$.
Another, closely related normalization for the radial wave function
sets its modulus squared equal to the rate of flow of energy.
We shall denote this choice by $\Psi$.  At large radii,
$\mathcal E \sim (\ddot h_+ + i \ddot h_\times) = \omega^2
(h_+ + i h_\times)$, where $h_+$ and
$h_\times$ are the dimensionless gravitational wave fields; so the
radiated energy is $\dot E_\infty \sim r^2 |\dot h_+ + i \dot h_\times|^2
\sim (r/\omega)^2 \mathcal E_\infty^2$, which tells us that $\Psi_\infty \sim
(r/\omega)\mathcal E_\infty$. Near the body's surface (i.e. near where the
horizon would be if the body were a Kerr black hole), the energy flux
is $\dot E \sim (r/\omega)^2 |\alpha^2 \mathcal E|^2$, where $\alpha$
is the Kerr-metric lapse function, which goes to zero at the horizon
radius.  (The ZAMOs' divergently large outward speed, relative
to infalling observers, causes them to see a divergently large tidal
field; the factor $\alpha^2$ corrects for that divergence; see, e.g.,
the discussion in Sec.\ VI.C.2 of \cite{mpbook}.)  Thus, in order to ensure
that the power flow is the square of the renormalized radial wave
function,
\begin{eqnarray}
\dot E \sim |\Psi|^2\;,
\label{eq:dotEPsi}
\end{eqnarray}
we must renormalize the ZAMO-measured tidal field $\mathcal E$ by
\begin{eqnarray}
\Psi \sim (r/\omega)\mathcal E \text{ at } r\rightarrow\infty\;,\nonumber\\ \;\;
\Psi \sim (\alpha^2 r/\omega) \mathcal E \text{ near body}\;.
\end{eqnarray}

\begin{figure}
\includegraphics[width=20pc]{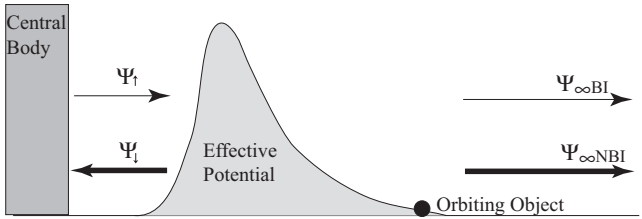}
\caption{The renormalized tidal gravitational fields $\Psi$ produced
near a central body's surface and at large radii
by the orbiting object, when the central body has the same exterior
metric as a Kerr black hole.}
\label{fig:Psi}
\end{figure}

With these preliminaries finished, we can give our physical argument
for Eq.\ (\ref{eq:Ratiovn})
in terms of the radial wave function $\Psi$.  Our argument relies on
Fig.\ \ref{fig:Psi}.

If the central body is a Kerr black hole, then the boundary condition
on $\Psi$ at its surface (the horizon) is purely downgoing waves, and
at infinity, purely outgoing waves.  The ratio of downgoing power
at the horizon to outgoing power at infinity has the standard 
Kerr values \cite{PoissonSasaki}\cite{TMT}:
$\dot E_\text{body}/\dot E_{\infty\text{NBI}} \sim v^8$ if the hole's 
spin angular 
velocity
$\Omega_H$ is much less than the orbital angular velocity $\Omega_\phi$;
and $\dot E_\text{body}/\dot E_{\infty\text{NBI}} \sim v^5$ if 
$\Omega_H \gg \Omega_\phi$.
(Here we have used the \emph{no-body-influence} notation 
$\dot E_{\infty\text{NBI}}$ 
for the
outgoing power because a central black hole's internal structure is unable to
influence the waves radiated to infinity.)
Correspondingly, by virtue of Eq.\ (\ref{eq:dotEPsi}), the ratio of the
downgoing field at the horizon $\Psi_{\downarrow}$ to the outgoing field at
infinity $\Psi_{\infty\text{NBI}}$ is
\begin{eqnarray}
\frac{\Psi_{\downarrow}}{\Psi_{\infty\text{NBI}}} \sim \left\{
\begin{array}{c} v^4 \\ v^{5/2}
\end{array}
\right\} \text{ for } \left\{
\begin{array}{l}  \Omega_H \ll \Omega_\phi \\ \Omega_H \gg \Omega_\phi
\end{array}
\right\} \;.
\label{eq:DownOverOut}
\end{eqnarray}
This suppression of the downgoing field relative to the outgoing is due,
mathematically, to a reflective effective potential in the wave equation
that $\Psi$ satisfies (Fig.\ \ref{fig:Psi}). Physically, it is due to
coupling of the field $\Psi$ to the central body's spacetime curvature.

Now suppose the central body is not a black hole, but some other object so
compact that its surface is well beneath the peak of the effective potential.  
This
mathematical assumption is equivalent to our physical assumption that the
ZAMOs see the downgoing field $\Psi_{\downarrow}$
so strongly blue shifted by the central body's gravity that it looks
like radiation.  The only way, then, that the central body can
influence the energy radiated to infinity is to reflect a portion
of this downgoing radiation back upward.  Mathematically, this corresponds
to replacing the black hole's downgoing boundary condition by
\begin{eqnarray}
\Psi_{\uparrow} = \mathcal R \psi_{\downarrow}
\label{eq:UpOverDown}
\end{eqnarray}
at some chosen radius just above the body's surface.  Here
$\Psi_{\downarrow}$ and $\Psi_{\uparrow}$
are the downgoing and upgoing components of $\Psi$; see Fig.\ \ref{fig:Psi}.
For simplicity, we shall assume that the amplitude reflection
coefficient $\mathcal R$ is small, $|\mathcal R|\ll 1$.
Otherwise we
would have to deal with a possible resonant buildup of energy between
the reflective central body and the reflective effective potential --- though
that would not change our final answer (see, e.g., the more detailed
analysis in the Appendix).

The upgoing waves $\Psi_{\uparrow}$ have great difficulty getting through
the effective potential. The fraction of the upgoing power that gets
transmitted through, successfully, is $\sim (M\omega)^6$ if the hole
rotates slowly, and $\sim (M\omega)^5$ if rapidly [Eq.\ (8.83) of
\cite{mpbook} with $\ell = 2$ and $\sigma_\infty = \omega$].  Since
the fields $\Psi$ are the square roots of the powers (aside from complex phase)
and since $M\omega = 2 M\Omega_\phi = 2 v^3$, this power
transmissivity corresponds to
\begin{eqnarray}
\frac{\Psi_{\infty\text{BI}}}{\Psi_{\uparrow}} \sim \left\{
\begin{array}{c} v^9 \\ v^{15/2}
\end{array}
\right\} \text{ for } \left\{
\begin{array}{l}  \Omega_H \ll \Omega_\phi \\ \Omega_H \gg \Omega_\phi
\end{array}
\right\}
\label{eq:OutBIOverUp}\;.
\end{eqnarray}

Combining Eqs.\ (\ref{eq:OutBIOverUp}), (\ref{eq:UpOverDown}), and
(\ref{eq:DownOverOut}), we see that
\begin{eqnarray}
\frac{\Psi_{\infty\text{BI}}}{\Psi_{\infty\text{NBI}}} \sim \left\{
\begin{array}{c} v^{13} \\ v^{10}
\end{array}
\right\} \text{ for } \left\{
\begin{array}{l}  \Omega_H \ll \Omega_\phi \\ \Omega_H \gg \Omega_\phi
\end{array}
\right\}\;.
\label{OutBIOverUp}
\end{eqnarray}

If these two complex outgoing fields are not precisely out of phase with
each other (phase difference $\pm \pi/2$), then the outgoing power is
$|\Psi_{\infty\text{NBI}} + \Psi_{\infty\text{BI}}|^2 \simeq
|\Psi_{\infty\text{NBI}}|^2 + 2 \Re \left(\Psi_{\infty\text{NBI}}
\Psi^*_{\infty\text{BI}}\right)$, which means that the ratio of
the radiated body-influenced power to radiated no-body-influence
power is
\begin{eqnarray}
\frac{\dot E_{\infty\text{BI}}}{\dot E_{\infty\text{NBI}}} \sim \left\{
\begin{array}{c} v^{13} \\ v^{10}
\end{array}
\right\} \text{ for } \left\{
\begin{array}{l}  \Omega_H \ll \Omega_\phi \\ \Omega_H \gg \Omega_\phi
\end{array}
\right\}\;.
\label{eq:BIoverNBIinfinity}
\end{eqnarray}
In the unlikely case (which we shall ignore) that the two fields are
precisely out of phase, the ratio will be the square of this.

By combining Eq.\ (\ref{eq:BIoverNBIinfinity}) with the square of
Eq.\ (\ref{eq:DownOverOut}), we obtain the ratio of the body-influence
power radiated to infinity over the tidal coupling power into the
central body:
\begin{eqnarray}
\dot E_{\infty\text{BI}}/\dot E_\text{body} \sim v^5
\label{eq:bodyOverBIinfinity}
\end{eqnarray}
independent of whether the body rotates slowly or rapidly.  This is
the claimed result.

If the central body's external metric is not Kerr, then the first-order
perturbation equations for the orbiting body's spacetime curvature will
probably not be separable in $\{r,\theta\}$, so the analysis will be
much more complex.  Nevertheless the physical situation presumably
is unchanged in this sense:  The body's spacetime curvature will couple to
the perturbation field in such a way as to resist energy flow through
the region between the body's surface and the object's orbit.  Correspondingly,
the perturbation fields and power flows are very likely to behave in the
same manner as for the Kerr metric, with the same final result,
$\dot E_\text{body}/\dot E_{\infty\text{BI}} \sim v^n$ with $n$ very
likely still 5 but possibly some other number significantly larger than one.

If this is, indeed, the case, then for any sufficiently compact central
body the power
tidally deposited into the body $\dot E_\text{body}$ will be very nearly
equal to
$\dot E_{\text{total},BI}$, which is measurable; and therefore
the tidal power will be measurable.

\section{Conclusion}
\label{sec:conclusion} In this paper, we have extended
Ryan's analysis to show that in principle it is possible to recover
not only the spacetime geometry of the central body,
but also the evolving orbital parameters of the inspiraling object
and the evolving
tidal coupling between the small object and the central body.
Therefore, in principle we can obtain a full description of the SARSAF
spacetime, the tidal coupling, and the inspiral orbit from EMRI or IMRI 
waveforms. 
In practice, the method of extracting the information is likely to be quite 
different 
from the algorithm 
we have presented here.

Further generalizations of Ryan's theorem and
development of practical methods to implement it are topics
of our ongoing research.

\begin{acknowledgments}
We would like to thank Yanbei Chen, Steve Drasco, Yi Pan, and 
Kip Thorne for helpful discussions, and Thorne for assistance with the prose 
of this 
paper. This work was supported in part 
by NSF grants PHY-0099568 and PHY-0601459, NASA grants NAG5-12834 
and NNG04GK98G, and the Brinson Foundation.
\end{acknowledgments}

\appendix

\section{An explicit derivation of results in Sec.~\ref{sec:ambiguity}}

\subsection{Teukolsky Perturbation Formalism}
In this subsection, we use the Teukolsky perturbation theory to
justify our results in Sec.~\ref{sec:ambiguity}. We first briefly
review the standard Teukolsky perturbation formalism. Details can
be found in e.g. Ref.~\cite{SasakiTagoshi}. To shorten our
expressions, in this appendix we restrict ourselves to a
nonrotating central body with external metric the same as a
Schwarzschild black hole but with a finite reflectivity.  The
generalization to the Kerr metric is straightforward but with more
cumbersome algebra.  We have carried it out, obtaining the same
result as is found by the physical argument in the text.

In the Teukolsky formalism, people usually calculate the
perturbation to a Newman-Penrose quantity $\psi_4$ that is related
to the ZAMO-measured tidal field $\mathscr{E}$ by a linear transformation of the
basis vectors. This $\psi_4$ can be decomposed into
Fourier-Harmonic components according to \begin{eqnarray}
\psi_4=\frac{1}{r^4}\int_{-\infty}^{\infty}d\omega\sum_{lm}R_{\omega
lm}(r){}_{-2}Y_{lm}(\theta,\phi)e^{-i\omega t},\end{eqnarray}
where ${}_{-2}Y_{lm}(\theta,\phi)$ are the spin-weighted spherical
harmonics. The radial function $R_{\omega lm(r)}$ satisfies the
inhomogeneous Teukolsky equation 
\begin{eqnarray} \left[r^2
\alpha^2\frac{d^2}{dr^2}-2(r-M)\frac{d}{dr}+U(r)\right]R_{\omega
lm}(r)=T_{\omega
lm},\label{eq:inhomogeneousequation}\end{eqnarray} where
$\alpha^2=1-2M/r$ is the lapse function for the Schwarzschild
metric. The expressions for the potential $U(r)$ and the source
$T_{\omega lm}$ can be found in e. g. , Ref.~\cite{PoissonSasaki},
Eqs. (2.3), (A1).

In order to solve this equation, we construct two linearly
independent solutions to the homogeneous Teukolsky equation, which
satisfy the following boundary conditions,
\begin{eqnarray}
R^\text{IN}_{\omega lm}&\hspace{-0.15cm}\rightarrow
\hspace{-0.15cm}&\left\{
\begin{aligned}&(\omega r)^4\alpha^4e^{-i\omega r^*},
&\hspace{-0.34cm} r\rightarrow 2M\\
&(\omega r)^{-1}Q^\text{in}_{\omega lm}e^{-i\omega r^*}+(\omega
r)^3Q^\text{out}_{\omega lm}e^{i\omega
r^*},&\hspace{-0.34cm}r\rightarrow +\infty
 \end{aligned}\right.\nonumber\\
 R^\text{UP}_{\omega
lm}&\hspace{-0.15cm}\rightarrow\hspace{-0.15cm}&\left\{
\begin{aligned} &(\omega r)^4\alpha^4 P^\text{out}_{\omega
lm}e^{-i\omega r^*}+P^\text{in}_{\omega lm}e^{i\omega
r^*},&\hspace{-0.35cm}
 r\rightarrow 2M\\
&(\omega r)^3 e^{i\omega r^*}, & \hspace{-0.35cm}r\rightarrow
+\infty
 \end{aligned}\right.\label{eq:stadinner}
 \end{eqnarray}
where $d/dr^*=\alpha^2 d/dr$. From these two homogeneous
solutions, we can construct the inhomogeneous solution according
to\begin{widetext}
\begin{eqnarray}
R_{\omega lm}(r) =\frac{1}{\text{Wronskian}[R_{\omega
lm}^\text{UP},R_{\omega lm}^\text{IN}]} \left( R_{\omega
lm}^\text{UP}(r) \int_{2M}^rdr'R_{\omega lm}^\text{IN}(r')
\mathcal T_{\omega lm}(r')+R_{\omega lm}^\text{IN}(r)
\int_r^\infty dr' R_{\omega lm}^\text{UP}(r') \mathcal T_{\omega
lm}(r')\right),\label{eq:inhomogeneoussolution}
\end{eqnarray}
\end{widetext}where $\mathcal T_{\omega lm}(r)\equiv
T_{\omega lm}(r)(r^2-2Mr)^{-2}$. This solution has only outgoing
waves at infinity and satisfies the \emph{purely ingoing boundary
condition:} (Ref.~\cite{PoissonSasaki}, Eqs.~(2.8) and
(2.9))\begin{eqnarray} R_{\omega lm}(r\rightarrow
\infty)&\sim& \mu \omega^2 Z^\text{IN}_{\omega lm}r^3e^{i\omega r^*},\nonumber\\
R_{\omega lm}(r\rightarrow 2M)&\sim&\mu\omega^3
Z^\text{UP}_{\omega lm}r^4\alpha^4 e^{-i\omega r^*},
\label{eq:zinzup}\end{eqnarray} where
\begin{eqnarray} Z^{\text{IN},\text{UP}}_{\omega lm}&=&\frac{1}{2i\mu \omega^2
Q^\text{in}_{\omega lm}}\\ &\times&\int_{2M}^\infty dr\left[
(r^2-2Mr)^{-2}R^{\text{IN},\text{UP}}_{\omega lm}(r)T_{\omega
lm}(r)\right].\nonumber\end{eqnarray} At infinity, where the
spacetime is almost flat, $\psi_4$ is directly related to the
outgoing gravitational wave strains according to
\begin{eqnarray}
\psi_4=\frac{1}{2}\left(\ddot{h}_+-i\ddot{h}_\times\right),\end{eqnarray}
and we can obtain the luminosity formula
(Ref.~\cite{PoissonSasaki}, Eq.~(2.21))\begin{eqnarray}
\dot{E}_\infty=\frac{1}{4\pi}\left(\frac{\mu}{M}\right)^2\sum_{lm}(M\omega)^2
|Z^\text{IN}_{\omega lm}|^2.\end{eqnarray}

\subsection{Inner Boundary Condition}
The above purely ingoing boundary condition makes sense when the
central body is a black hole because we know everything is
absorbed at the horizon of the black hole. If the central body is
some other kind of object, the only way it can influence the
perturbation field $R_{\omega l m}$ just above its surface is by
producing an outgoing-wave component via some effective
reflectivity $\mathcal R$. The result will be a modified field
\begin{eqnarray} R_{\omega lm}(r\rightarrow 2M)\sim
e^{-i\omega r^*}+(\text{something}) e^{i\omega r^*}.\end{eqnarray}
The ``something'' will be proportional to $\mathcal R$, and it
will also have a peculiar radial dependence because $\psi_4$
relies for its definition on an ingoing null tetrad and thereby
treats ingoing and outgoing waves in very different manners.
\subsection{Chandrasekhar Transform}
To learn what the ``something'' should be, we can transform to a
new radial wave function that treats ingoing and outgoing waves on
the same footing.  Two such functions were introduced and used in
Sec.~\ref{sec:ambiguity}: the ZAMO-measured tidal field $\mathcal
E$ and a field $\Psi$ whose modulus squared is the power flow, for
both outgoing and ingoing waves. Those choices are good for
Sec.\ref{sec:ambiguity}'s physical, order-of-magnitude arguments,
but at general radii $r$ they not related in any simple way to
$\psi_4$. A choice that \emph{is} simply related to $\psi_4$ is
the Regge-Wheeler function $X$, and we shall use it here.

The radial wave function $R$ for the Newman-Penrose $\psi_4$ is
related to the Regge-Wheeler function $X$ by the Chandrasekhar
transform, Eq.~(A6) of Ref.~\cite{PoissonSasaki}.  This Chandrasekhar
transform takes the form
\begin{eqnarray}R_{\omega lm}^{\text{IN},\text{UP}}=\chi_{\omega
lm}^{\text{IN},\text{UP}}C_\omega X^{\text{IN},\text{UP}}_{\omega
lm},\end{eqnarray} where
\begin{eqnarray}\chi_{\omega lm}^\text{IN}&=
&\frac{16(1-2iM\omega)(1-4iM\omega)(1+4iM\omega)}{(l-1)l(l+1)(l+2)-12iM\omega}
(M\omega)^3,
\nonumber\\\chi_{\omega
lm}^\text{UP}&=&-\frac{1}{4}.\label{chieq}\end{eqnarray}$C_\omega$
is a second order differential operator, and $X_{\omega
lm}^{\text{IN},\text{UP}}$ are two linearly-independent solutions
of the homogeneous Regge-Wheeler equation
\begin{eqnarray}
\left[\frac{d^2}{dr^{*2}}+\omega^2-V(r)\right]X_{\omega
lm}(r)=0\label{eq:regge},\end{eqnarray} where\begin{eqnarray}
V(r)=\alpha^2\left[\frac{l(l-1)}{r^2} -\frac{6M}{r^3}\right].
\end{eqnarray}

The asymptotic expressions for $X_{\omega
lm}^{\text{IN},\text{UP}}$ are (Ref.~\cite{PoissonSasaki},
Eq.~(2.7)) \begin{eqnarray} X^\text{IN}_{\omega
lm}&\rightarrow&\left\{
\begin{aligned} & e^{-i\omega r^*},& r\rightarrow 2M\\
&A^\text{in}_{\omega lm}e^{-i\omega r^*}+A^\text{out}_{\omega
lm}e^{i\omega r^*},& r\rightarrow +\infty
 \end{aligned}\right.\\
 X^\text{UP}_{\omega lm}&\rightarrow
 &\left\{ \begin{aligned} &-B^\text{out}_{\omega lm}e^{-i\omega r^*}+
B^\text{in}_{\omega lm}e^{i\omega r^*},& r\rightarrow 2M\\
&e^{i\omega r^*}, & r\rightarrow +\infty
 \end{aligned}\right.\nonumber\label{eq:Xbound}\end{eqnarray}
Here we note that by the conservation of the Wronskian, it is
straightforward to show that $B^\text{in,out}=A^\text{in,out}$.

\subsection{$\dot{E}_\infty$ with a Reflective Inner Boundary Condition }
Because the Regge-Wheeler function treats outgoing and ingoing
waves on the same footing, the desired, reflective inner boundary
condition for it takes the simple form
 \begin{eqnarray} \tilde{X}^\text{IN}_{\omega lm}(r\rightarrow 2 M)&\sim&
e^{-i\omega r^*}+\mathcal{R} e^{i\omega
r^*}.\label{eq:horizonbound}\end{eqnarray} Here
$\tilde{X}^\text{IN}_{\omega lm}$ is a new homogeneous solution of
the Regge-Wheeler equation.

This new homogeneous solution is a superposition of both ingoing
and outgoing waves at the horizon. It is shown in
Ref.~\cite{Chandra} that because the Regge-Wheeler function treats
outgoing and ingoing waves in the same manner, $|\mathcal{R}|^2$
has the physical meaning of the energy flux reflectivity, i.e.,
the ratio between outgoing and ingoing energy flux at the
horizon.

The homogeneous solution (\ref{eq:horizonbound}) which satisfies
the new inner boundary condition can be constructed from the old
homogeneous solutions: \begin{eqnarray}
\tilde{X}^\text{IN}_{\omega lm}=\beta_1 X^\text{IN}_{\omega
lm}+\beta_2 X^\text{UP}_{\omega lm},\end{eqnarray} where
\begin{eqnarray} \beta_1=1
+\frac{\mathcal{R} A^\text{out}_{\omega lm}}{A^\text{in}_{\omega lm}},
\quad \quad
\beta_2=\frac{\mathcal{R}}{A^\text{in}_{\omega
lm}}.\label{eq:beta1beta2}\end{eqnarray}

After doing an inverse Chandrasekhar transform, we obtain the
corresponding homogeneous solution of the homogeneous Teukolsky
equation\begin{eqnarray} \tilde{R}^\text{IN}_{\omega
lm}=R^\text{IN}_{\omega
lm}+\frac{\beta_2}{\beta_1}\frac{\chi^\text{IN}_{\omega
lm}}{\chi^\text{UP}_{\omega lm}}R^\text{UP}_{\omega
lm}.\end{eqnarray}

Now we can replace $R^\text{IN}$ by $\tilde{R}^\text{IN}$ in
Eq.~\eqref{eq:inhomogeneoussolution} to obtain the solution
$\tilde R_{\omega lm}(r)$ which satisfies the inhomogeneous
Teukolsky equation with upgoing and downgoing waves at the
horizon and purely outgoing waves at infinity. From this $\tilde
R_{\omega lm}(r)$ we identify the new amplitudes $\tilde
Z^\text{IN}_{\omega l m}$ as in Eq.~\eqref{eq:zinzup}:
\begin{eqnarray} \tilde{Z}^\text{IN}_{\omega
lm}=Z^\text{IN}_{\omega
lm}+\frac{\beta_2}{\beta_1}\frac{\chi_{\omega
lm}^\text{IN}}{\chi_{\omega lm}^\text{UP}}Z^\text{UP}_{\omega
lm}.\end{eqnarray} From these new $\tilde Z^\text{IN}_{\omega l
m}$ the calculation of the luminosity at infinity is
straightforward.

In Ref.~\cite{PoissonSasaki} Poisson and Sasaki have already
worked out all the relevant formulae, so we only give the
results. For the original expressions in
Ref.~\cite{PoissonSasaki}, please refer to Eq.~(3.25) for
$A^\text{in},A^\text{out}$, Eq.~(A7) for
$\chi^\text{IN},\chi^\text{UP}$, Eqs.~(5.4), (5.6), (5.11), (5.12)
for $Z_{\omega lm}^\text{IN},Z_{\omega lm}^\text{UP}$.

The leading luminosity correction comes from the $l=2,m=\pm 2$
mode,
and we have
\begin{eqnarray}
\left.\dot{E}\right._\infty=\left.\dot{E}_\infty
\right|_\text{Schwarzschild}\left|1-\frac{128i\mathcal{R}
v^{13}}{15\beta_1}\right|^2,
\end{eqnarray}
where $v$ is the same PN expansion parameter
as that in Sec.~\ref{sec:ambiguity}. Unless the reflection
coefficient $\mathcal{R}$ is precisely real, this gives
\begin{eqnarray}
\left.\dot{E}\right._\infty=\left.\dot{E}_\infty
\right|_\text{Schwarzschild}
\left[1+\frac{256}{15}\Im\left(\frac{\mathcal{R}}{\beta_1}\right)
v^{13}\right]
\end{eqnarray}
in agreement with Eq.\ (\ref{eq:BIoverNBIinfinity}). The change
in $\dot{E}_\text{body}$ should be \begin{eqnarray}
\left.\dot{E}_{\text{body}}\right.=\left.\dot{E}_\text{body}
\right|_\text{Schwarzschild}\left(\frac{1-|\mathcal{R}|^2}
{|\beta_1|^2}\right)\label{newhole}\end{eqnarray} where $\beta_1$
is defined in Eq.~\eqref{eq:beta1beta2}.


\end{document}